# Recognising natural capital on the balance sheet: options for water utilities

Marie-Chantale Pelletier[a], Claire Horner[b,c], Mathew Vickers[a], Aliya Gul[d], Eren Turak[d], Christine Turner[e]

a. Faculty of Science and Engineering, Southern Cross University, Military Road, East Lismore, NSW
b. CSIRO, 15 College Road, Sandy Bay, Tasmania
c. Integrated Futures, Canberra, ACT
d. NSW Department of Planning and Environment, 4 Parramatta Square, Parramatta, NSW
e. Sydney Water Corporation, 1 Smith St, Parramatta, NSW

## Abstract

**Purpose**

The aim of this study was to explore the feasibility of natural capital accounting for the purpose of strengthening sustainability claims by reporting entities. The study linked riparian land improvement to ecosystem services and tested options for incorporating natural capital concepts into financial accounting practices, specifically on the balance sheet.

**Methodology**

To test the approach, the study used a public asset manager (the water utility) with accountabilities to protect the environment including maintaining and enhancing riparian land assets.  Research activities included stakeholder engagement, physical asset measurement, monetary valuation and financial recognition of natural capital income and assets. Natural capital income was estimated by modelling and valuing ecosystem services relating to stormwater filtration and carbon storage.

**Findings**

This research described how a water utility could disclose changes in the natural capital assets they manage either through voluntary disclosures, in notes to the financial statements or as balance sheet items. We found that current accounting standards allowed the recognition of some types of environmental income and assets where ecosystem services were associated with cost savings. The proof-of-concept employed to estimate environmental income through ecosystem service modelling proved useful to strengthen sustainability claims or report financial returns on natural capital investment.

**Originality/value**

This study applied financial accounting processes and principles to a realistic public asset management scenario with direct participation by the asset manager working together with academic researchers and a sub-national government natural resource management agency. Importantly it established that natural assets could be included in financial statements, proposing a new approach to measuring and reporting on natural capital.

## 1. Introduction

There is a growing international consensus that recognising natural capital assets alongside other economic assets can steer investment away from negative outcomes and towards nature-positive outcomes. This consensus was recently validated by the Independent Review on the Economics of Biodiversity commissioned by the UK Treasury (Dasgupta, 2021), which clearly articulated the need to go beyond sustainable management and increase nature's supply, to change the way we measure success by accounting for natural capital assets and to transform financial institutions to channel investment towards natural capital and mitigate risk resulting from unsustainable management.

There has been significant progress in the development of accounting frameworks to assist such a transition. The System of Environmental-Economic Accounting (SEEA) provides robust internationally accepted concepts and methods to organise information linking nature to economic statistics. It builds on national accounting practices to expose the dependency, but also the impacts, of the economy on natural capital (UN, 2023a). The SEEA Central Framework provides advice for specific aspects of natural capital such as energy, land, water, air emissions, and agriculture. By contrast, the SEEA Ecosystem Accounting (SEEA-EA) describes natural capital as ecosystems providing services that generate benefits for people (UN *et al*., 2021). Like other economic assets, natural capital assets need to be maintained over time and investment is required to do so.

Government initiatives around the world are working to recognise nature's contribution to wellbeing more explicitly in financial reporting. Estimating the value of natural capital using an accounting methodology such as the SEAA Framework is being used to enhance sustainability practices and assist in decision making to address broader challenges including climate change and biodiversity decline. In the UK the Office of National Statistics has been publishing Environmental Accounts since the 1990s and experimental Natural Capital Accounts since 2014 (ONS, 2023a; ONS 2023b). Other leading agencies in Canada, Australia and the European Union are collaborating to progress methods and overcome challenges (Hein *et al*., 2020; Bagstad *et al*., 2021; Dasgupta, 2021); Chen *et al*., 2023). Some international initiatives are also incorporating the accounting framework for monitoring and reporting progress on environmental outcomes. The UN Convention on Biological Diversity (CBD) Kunming – Montreal Global Biodiversity Framework includes 2030 targets for the valuation of nature through green investments, ecosystem service valuation in national accounts, and public and private sector financial disclosure (CBD, 2022)

Alongside these government initiatives, an increasing number of organisations are voluntarily reporting on their sustainability performance (KPMG, 2022), including government and non-government organisations. Sustainability reporting through voluntary disclosures may have helped grow awareness of the need to recognise environmental impacts and dependencies, however it has not been successful in achieving substantive corporate disclosures (Davern *et al*., 2021). Rather, the increase in corporate sustainability reporting has seen a corresponding increase in the provision of misleading information, or conveying a false impression to mislead consumers or investors about the environmental friendliness of the organisation's products or activities (Hsiao *et al*., 2022).

Guidance on methods and metrics to improve voluntary disclosure include the Global Reporting Initiative (GRI), Sustainability Accounting Standards Board (SASB), Natural Capital Protocol (NCC, 2016) and ESG metrics recommended by the World Economic Forum's International Business Council (World Economic Forum, 2020). Efforts are growing to recognise the dependency of business on natural capital as well as the impacts of business on nature (Capitals Coalition, 2016; Ingram *et al.*, 2022), the purpose being to improve financial performance, manage risk associated with natural assets and meet stakeholder expectations (World Economic Forum, 2020). However, while this guidance provides some options for disclosing impacts and dependencies on nature, it has not led to the level of commitment and legal obligations associated with assets as recognised under International Accounting Standards (IASs).

There have been recent developments in the mandatory sustainability reporting environment, with the International Sustainability Standards Board (ISSB) issuing IFRS S1 *General Requirements for Disclosure of Sustainability-related Financial Information* and IFRS S2 *Climate-related Disclosures.* These Standards incorporate the recommendations of the Task Force on Climate-related Financial Disclosures (TCFD), the role of which has since been fully subsumed by the IFRS Foundation (IFRS Foundation, 2023). Whether IFRS S1 and S2 become mandatory depends on whether individual jurisdictions choose to legislate them, and a number of countries, including the UK and Australia, have indicated that they will issue their own Standards based on S1 & S2, with the Australian Accounting Standards Board (AASB) issuing their Exposure Draft in October 2023.

The Australian Accounting Standards Board (AASB) is an Australian Government agency that is responsible for developing a conceptual framework to evaluate proposed accounting standards, create new accounting standards, and participate in and contribute to the development of a single set of accounting standards for use worldwide. The AASB enables Australian entities to effectively compete in domestic and overseas capital markets and to maintain investor confidence in the Australian economy (AASB, 2023). All Commonwealth, State and Territory entities are required to prepare financial statements in accordance with the Australian Accounting Standards (AAS) (Department of Finance, 2021), and the conceptual framework may provide guidance to reporting entities in the absence of specific instruction in the AAS.

The sister initiative of the TCFD, the Task Force on Nature-related Financial Disclosures (TNFD) released its final recommendations for disclosure requirements in September 2023. Similar to IFRS S1 and S2, the recommended disclosures are structured around governance, strategy, risk and impact management, as well as metrics and targets (TNFD, 2023). Assuming that the TNFD follows the same path as the TCFD, these recommendations will form the basis of IFRS S3. However, the TNFD Framework has been criticised by environmental groups for failing to take a double-materiality approach, by only focusing on the financial risks to the business, rather than also considering the risks the business poses to nature, and potentially facilitating corporate greenwashing (Hawkes, 2022; Sutherlin, 2023). One way of mitigating against the risk of greenwash and to strengthen sustainability reporting is a robust accounting system (Horner, 2014), and the recognition and measurement of natural capital can be used as a means to achieve positive impacts (CPA, 2022).

Therefore, the purpose of this study is to explore whether natural capital concepts can be incorporated into existing financial accounting practices. Specifically, we ask: Can natural capital be recognised in financial statements under existing financial accounting Standards? And if so, what methods can be used in a water utility to recognise natural capital on the balance sheet? In order to answer these questions, we collaborated with Sydney Water, a large water utility in New South Wales (NSW), Australia. An overview of Sydney Water and their operations is provided in the following section. The remainder of the paper is structured as follows: Section 2 describes the agency managing the assets used in the research, the conceptual background is provided in Section 3, the research methods are discussed in Section 4 and the findings are presented in Section 5.

## 2. A practical application in a water utility

As noted above, we collaborated with Sydney Water to demonstrate whether and how an entity can incorporate public natural capital assets into their financial reporting. As the extent of responsibilities of water utilities varies widely, these are often shared with other entities through co-management arrangements. In Australia, these legislative arrangements can span local, state and federal government jurisdictions. Responsibilities include a mix of water supply, wastewater treatment and stormwater management. The size of the catchment, the range of land uses present in the catchment and the rate of land use change all vary widely among utility organisations. Rural water utilities for example tend to experience less rapid change than urban ones, but their distribution networks tend to be longer and the number of customers lower.

The Australian water utility Sydney Water is a sub-national government-owned statutory corporation providing drinking water and wastewater services to over 5 million people in Sydney and surrounds (Sydney Water, 2022). The utility also provides recycled water and stormwater services to specific areas including the study area at Rouse Hill, a sub-catchment of the Hawkesbury-Nepean River system located about 45 km northwest of the Sydney CBD. The focus for this 'proof of concept' project was on the Rouse Hill riparian land assets and specific flood zones, most of which are owned and/or managed by Sydney Water.

For the purpose of this study, we consider a subset of Sydney Water's responsibilities relating to stormwater management in an urban setting. Most water utilities have objectives around environmental protection and improved valuation of environmental resources that contribute to their business objectives. The naturalisation and revitalisation of stormwater assets are an example of natural capital assets that improve the quality of water released to waterways. These 'soft engineered' approaches typically comprise natural or constructed wetlands for sediment control and vegetated riparian land assets that filter overland precipitation runoff before it enters adjacent streams. More dense vegetation in the riparian zone generally performs better than sparse vegetation, bare land or impervious surfaces such as roads and building roofs. In addition to water filtration, natural assets provide extra benefits to the community including climate mitigation in the form of carbon capture and sequestration, habitat for biodiversity, protection against soil erosion, flood mitigation and recreation opportunities.

Maintaining and improving vegetation in the riparian zone to meet both stormwater quality standards as well as community expectations of environmental stewardship requires adequate funding. Sufficient funding can be difficult to secure as the benefits provided by natural assets are not easily valued in standard accounting practices, and consequently their contribution to business activities and social wellbeing may be underestimated in investment decisions. Natural assets are generally not recognised the same way grey infrastructure assets are, despite delivering similar benefits. As a result, improving riparian land vegetation may not be easily identified as an essential activity requiring ongoing funding. This under-representation as fully-fledged assets can lead to inconsistencies in performance evaluation of natural assets relative to grey infrastructure.

A natural capital accounting approach that recognises a water utility's natural assets provides a more systematic assessment for direct input into operational and strategic decision making. This is important in decision making as it can assist in steering appropriate investment to natural stormwater assets to achieve multiple sustainability objectives in new, urbanising catchments, as in the Rouse Hill sub-catchment. Natural capital accounting therefore has direct application into business strategic planning, investment decisions, supply chain management, operations management, risk management, and corporate reporting (Ingram *et al.,* 2022) which are important business management processes in the water industry. It is also important for demonstrating responsible stewardship over the natural environment under the water utility's control, in order to meet stakeholder expectations as well as the proposed reporting requirements of the TNFD (2023) and other nature-related commitments.

## 3. Conceptual background

In this Section we first introduce the SEEA-Ecosystem Accounting Framework, followed by an exploration of the extant literature applying the SEEA Framework to water utilities, and a discussion of the implications of this for environmental stewardship, which together form the conceptual and theoretical framework for this study.

### 3.1 The SEEA-Ecosystem Accounting

This study drew from existing frameworks supporting the recognition of natural capital, combining specific elements best suited to the context of public sector agencies. Figure 1 shows the contribution of the SEEA-EA, the Natural Capital Protocol and the Australian Accounting Standards frameworks to different stages of the study.

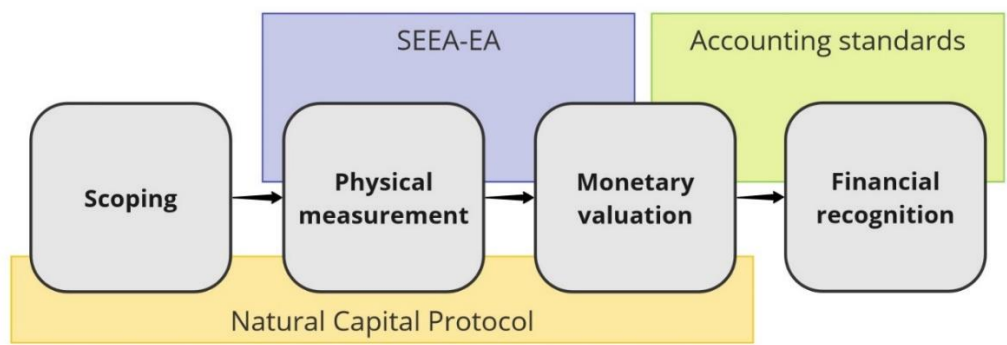

Figure 1. Steps in the research process and relevance of the accounting frameworks supporting the study. Source: Produced by the authors.

The Natural Capital Protocol (the Protocol) is a decision-making framework that enables organisations to identify, measure and value their direct and indirect impacts and dependencies on natural capital (Capitals Coalition, 2016). The first step in accounting for natural capital consists of reporting on the assets owned and controlled by the reporting agency. This process allows an agency to develop internal capacity and to establish relevant information systems.

The SEEA-EA provides accounting concepts and methods to strengthen the measurement and valuation of natural capital. Ecosystems are conceived as assets with extent and condition characteristics that determine their capacity to provide multiple goods and services to beneficiaries (UN *et al.*, 2021). For example, a forested land parcel (asset) may be 100 hectares in size (extent), have low species diversity (against benchmark condition) and provide carbon sequestration benefits (service) to the global community. A wetland (asset) may cover 1 hectare in size (extent), have high turbidity (condition) and provide flood mitigation benefits (service) to local residents. A given ecosystem asset may provide multiple services simultaneously to diverse stakeholders, though it may not always be reportable: the forest that sequesters carbon for the landholder may also provide visual amenity for nearby residents, or a pollinator habitat for other nearby systems, agricultural or natural. In this case, the landholder may not be able to account directly for the habitat provision service being provided.

The concept of ecosystem services aligns with the concept of income streams which require physical measurement and monetary valuation. Asset valuation is achieved either using market prices or the net present value approach already used in financial accounting. Figure 2 shows the relationship between the main elements of ecosystem accounting. These concepts and methods, described in detail in the SEEA-EA standard, provide a robust measurement framework to support natural capital accounting.

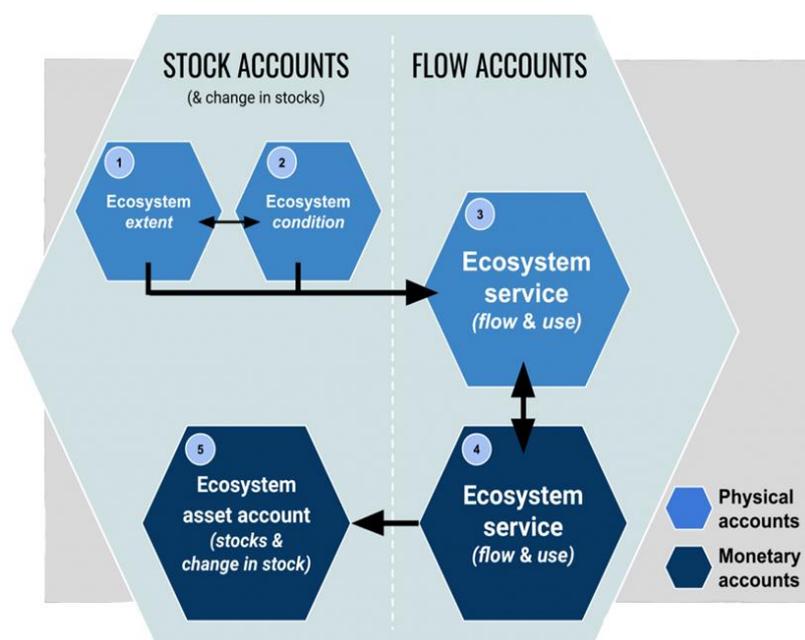

Figure 2.  Stock and flow accounts described in the SEEA-EA Framework. Source: United Nations, 2023b.

Although the SEEA was initially designed to align with national accounting practices for reporting on national territories, many of its elements can be applied to subsets of national ecosystems managed by government agencies or by private entities in a manner that helps meet their objectives and aligns with their corporate accounting practices (Bagstad et al., 2021; Ingram et al., 2022). While a general awareness of the relevance of the SEEA to business accounting is growing, there have been very few examples so far demonstrating practical applications of the approach. SEEA methods for valuing the contribution of ecosystems to grazing enterprises have been found to be consistent with Australian Accounting Standards (Ogilvy and Vail, 2018). Ecosystem degradation resulting from overuse by pastoral activities could also lead to the recognition of financial liabilities under current international accounting standards, where there is an expectation the ecosystem condition would not be compromised by commercial activities (Ogilvy *et al.*, 2018). Without explicitly referencing the SEEA framework, Houdet et al. (2020) demonstrated how biophysical measures of biodiversity could be recorded in Statements of Biodiversity Performance and Position to report the physical (but not financial) impacts of business on natural capital.

## 3.2 Applications of the SEEA to water resources

The majority of studies exploring the application of the SEEA to water management have been applied to national or sub-national territories, reporting on water assets from a natural resource perspective (SEEA Central Framework) with the view to inform government policy and sustainability reporting (Remme *et al.*, 2015; Salminen *et al.*, 2018; Mahdavi *et al.,* 2019; Torres-Lopez *et al.,* 2019; Esen and Hein, 2020; Bagstad *et al.,* 2020; Esen and Hein, 2020; Esmail *et al.,* 2023). Very little research has taken a broader ecosystem perspective to consider the role of vegetation in filtering water before it enters water bodies, which is a regulating ecosystem service (Warnell *et al.,* 2020; Bagstad *et al.,* 2020; Esmail *et al.,* 2023; Boschetto *et al.,* 2023). Most of these studies estimated the physical flow of services without extending to monetary valuation. Bagstad et al. (2020) report the value of water supplied by ecosystems in the US but only report physical flows of filtration services. Boschetto et al. (2023) consider instead the water retention and storage capacity of soil as a function of soil characteristics and land cover to describe the water provisioning service, including their monetary values. The effect of soils and land cover in retaining sediments and nutrients however is not assessed.

One study in southern Victoria, Australia showed how an ecosystem service perspective can be used to inform the design of natural assets managed by urban authorities and water utilities, although the work did not extend to incorporating the information into financial accounts (Ghofrani *et al.* 2020). Quantifying and valuing stormwater abatement, water quality improvement and water provisioning services was found to be useful in highlighting the multifunctional nature of natural capital in urban blue-green infrastructure. Using catchments in the Central Highlands of Victoria, Australia, Vardon et al. (2019) describe the SEEA-EA accounting treatment of water provision and filtration for abstracted water supplied to households and businesses in Melbourne. The authors take a national accounting perspective and describe alternatives for reporting flows of services, including between ecosystems, for example between a forest and water reservoir. They discuss the role of vegetation cover in filtering water supplies and describe how water filtration services could be reported alongside water provisioning services. However, due to a lack of available information, they report quantities of water filtered rather than quantities of sediments and pollutants avoided. The study further estimates monetary values for water provisioning services using a replacement cost method, but omits monetary values for the filtration services.

The present study extends previous work from multiple perspectives:

a) We quantify the change in water filtration service according to changes in the condition of vegetation cover;
b) We estimate the value of the filtration services provided by vegetation in monetary terms
c) We consider stormwater assets rather than water supply assets, accounting for water filtration alone, not water provisioning.
d) We take a corporate, rather than national accounting perspective, reporting the level of filtration services provided by a subset of assets controlled by Sydney Water, rather than reporting for a whole territory controlled by government.

To the best of our knowledge, there is no published literature explicitly linking information on the physical condition of terrestrial ecosystems to financial accounts in the context of water utilities. Doing so has important implications for the discharge of accountability from an environmental stewardship perspective. Environmental stewardship is defined as "the wise and responsible use of natural resources [to] support social-ecological resilience and human well-being" (West *et al.*, 2018, p. 30). The determination of what could be considered 'wise and responsible' has evolved to take on a range on meanings (Horner and Davidson, 2020; Mathevet *et al.*, 2018), however, according to the Australian Government's Strategy for Nature, "stewardship of nature can contribute to Australia's nature conservation objectives and also build the health and resilience of our society, businesses and economy" (Commonwealth of Australia, 2017, p. 10).

Sydney Water recognises the importance of their environmental stewardship, in that their operations traverse environmentally sensitive areas, including threatened ecological communities (Sydney Water, 2021). Environmental stewardship has often been used as a theoretical underpinning for studies seeking to operationalise environmental accounting and management (e.g. Horner and Davidson (2020); Hossain (2017); Jones (2003); Lobley et al. (2013)), and provides a useful framework for interdisciplinary projects such as this, as it facilitates collaboration between researchers and non-researchers of different backgrounds on areas of shared concern (West *et al*., 2018).

## 4. Research methods

The research methods used in this study followed the steps identified in Figure 3, which are discussed in greater detail below.

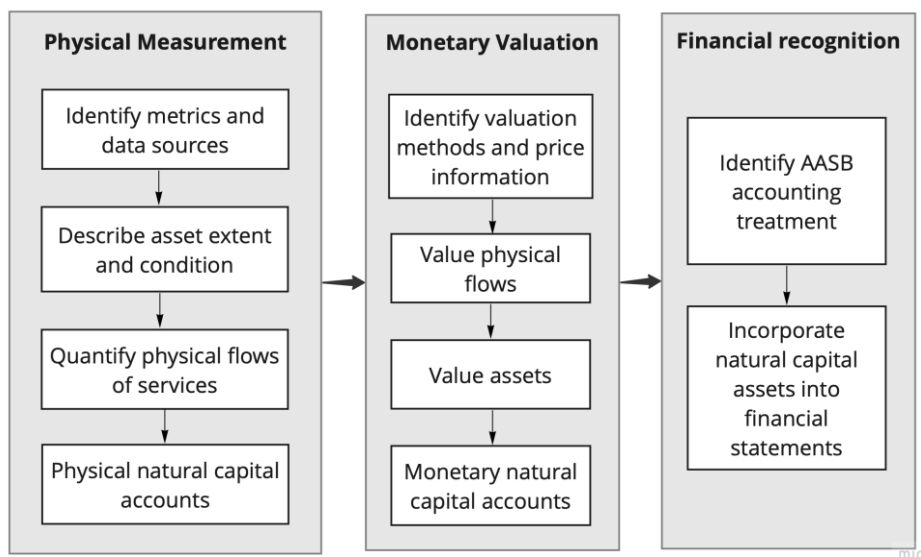

Figure 3. Steps in the process of recognising natural capital assets explored in the research. Source: Produced by the authors.

## 4.1 Asset extent

We selected the Rouse Hill riparian lands assets which provide water filtration services for the increasing stormwater flows from urban growth. The riparian land assets were classified according to the nature of the land cover present at the time of survey (Figure 4). The spatial data used for this study was collated from various publicly available sources: SRTM-derived 1 Second Digital Elevation Models from Geoscience Australia (Gallant *et al*., 2011), vegetation cover from the NSW State Vegetation Type Map (DPE, 2022), land use from Sixmaps (NSW Department of Customer Service - Spatial Services, 2022). The area consisted of five native plant Community types, Grass, Non-vegetated still water body, Watercourse and Medium Density Urban Fabric (Figure 4).

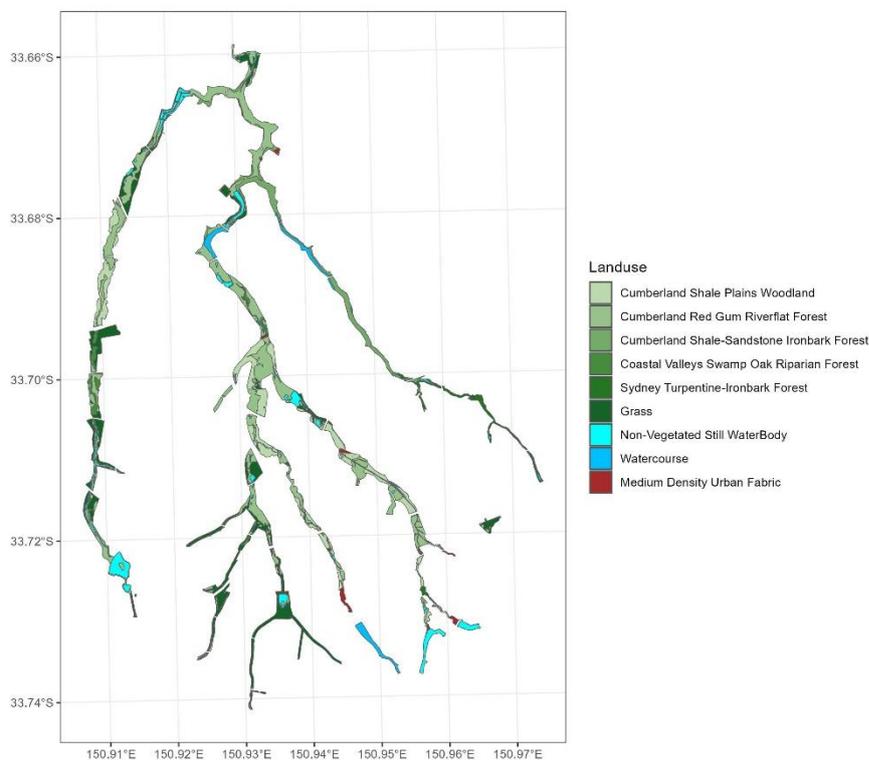

Figure 4. Map of riparian land assets used in the study. Source: Produced by the authors.

Following the SEEA-EA guidance, the extent of the riparian vegetation (Table 1) refers to the spatial area occupied by ecosystem assets, in this case riparian land assets (United Nations *et al*., 2021). Asset extent is described in hectares or square kilometres at the beginning and closing of the accounting period. The total area of riparian land assets in this example was 369 hectares.

Most environmental change, including incremental natural regeneration may not be meaningful on annual cycles. Environmental accounting therefore tends to span longer time periods in order to capture meaningful change. The accounting period for this study spans 2013 to 2023. We assumed the water utility did not acquire or sell any riparian land assets over the accounting period, so no change in asset extent was recorded. Ecosystem extent was assessed using spatial analysis carried out using free and open source QGIS and R software (R Core Team, 2023; QGIS Development Team, 2023).

Table 1. Asset extent of the ecosystem types scored in this study in hectares. Source: Produced by the authors.

| Asset type | Opening balance 2013 | Additions | Losses | Closing balance 2023 | Change |
|---|---|---|---|---|---|
| Cumberland Shale Plains Woodland | 58 | 0 | 0 | 58 | 0 |
| Cumberland Red Gum Riverflat Forest | 141 | 0 | 0 | 141 | 0 |
| Cumberland Shale-Sandstone Ironbark Forest | 26 | 0 | 0 | 26 | 0 |
| Coastal Valleys Swamp Oak Riparian Forest | 9 | 0 | 0 | 9 | 0 |
| Sydney Turpentine Ironbark Forest | 9 | 0 | 0 | 9 | 0 |
| Grass | 68 | 0 | 0 | 68 | 0 |
| Non-vegetated still waterbody | 39 | 0 | 0 | 39 | 0 |
| Watercourse | 12 | 0 | 0 | 12 | 0 |
| Medium Density Urban Fabric | 7 | 0 | 0 | 7 | 0 |
| Total | 369 | 0 | 0 | 369 | 0 |

## 4.2 Measurement of physical ecosystem services

Ecosystem services are the contributions of nature to economic production and social wellbeing. They correspond to the accounting concept of economic benefits flowing or expected to flow to the controlling entity (AASB, 2019). A measure of the physical quantity of ecosystem services flows has multiple uses for a water utility. It can be used by internal managers as a measure of asset performance to support budget allocation decisions. It can also be used to report on investment performance or to predict the outcomes of alternative investment, or to communicate environmental stewardship to external stakeholders. Here, we focus on stormwater filtration services and carbon services relating to global climate regulation. These were identified in the study scoping stage as being core business objectives for the water utility. They do not span the whole range of benefits provided by the assets, but they provide the most robust measurement opportunity. We compared ecosystem services in 2013 with a theoretical "optimal" scenario in 2023 ('the optimal 2023 scenario').

Ecosystem service performance varies through time but also in space across an area of interest, for instance, some areas sequester carbon faster than others. We capitalised on this spatial variability and simulated the optimal 2023 scenario where asset management activities such as tree planting, fertiliser use and weed control led to higher condition riparian land assets by 2023. This comparison of 2013 with the optimal 2023 scenario illustrates the impact of a new land management scenario without confounding effects of land use changes across the entire catchment, for example new urban development farther up in the catchment.

### 4.2.1 The optimal 2023 scenario

The optimal 2023 scenario relied upon the same spatial data, except for one parameter each in the sediment filtration and in the carbon storage models. In the sediment filtration model, the maximum filtration performance observed for native vegetation in 2013 was used to inform filtration performance for native vegetation in the optimal 2023 scenario. This means we assumed that management actions were taken by the water utility to improve the vegetation in the riparian land assets with a view to optimising water filtration services, while other land within the catchment remained constant. In the carbon storage model, maximum carbon storage observed in the native vegetation of 2013 was used to inform carbon storage for the optimal 2023 scenario. This assumes that management actions between 2013 and 2023 optimised carbon storage.

### 4.2.2 Sediment filtration

The InVEST (Integrated Valuation of Ecosystem Services and Tradeoffs) Sediment Delivery Ratio (SDR) model (version 3.11.0, Sharp *et al*., 2018) was used to estimate sediment trapped by vegetation within the catchment. Waterflow is estimated across the watershed interest based on a hydrologically-enforced digital elevation map and annual rainfall data. This, along with soil properties, allows calculation of the rate of sediment generation, runoff and entrapment for every pixel as tonnes/hectare using the Revised Universal Soil Loss Equation (RUSLE). The RUSLE is a widely used model (Sharp *et al*., 2018) that predicts soil erosion caused by overland flows. The RUSLE model incorporates average annual rainfall, soil erodibility, the likeliness of soil to erode, the length and steepness of the slope, the crop or vegetation cover, and conservation practices/landscaping for the area. The InVEST implementation estimates sediment loss, transport, and trapping, per pixel, which is summarised through aggregate statistics. Data used to parameterise the RUSLE model, rainfall, erosivity, and erodibility factor, were extracted from a comprehensive spatial model of hillslope erosion in NSW (Yang, 2014), other parameters were informed by recent literature (Bakker *et al*., 2008, Kouli *et al*., 2009).

### 4.2.3 Carbon storage

Carbon sequestration includes both additions (plant growth) and losses (respiration, fires, etc.), and corresponds to a flow of carbon from the atmosphere to the ecosystem asset over some period, in this case 10 years. Net sequestration can be positive (adding to the carbon pool) or it can be negative (emitting carbon from the pool). Using remote sensing data, carbon sequestration can be estimated as the difference in carbon storage among contiguous years. The InVEST carbon model (Sharp *et al*., 2018) used three carbon pools, above-ground, below ground, and dead-carbon pools, to estimate carbon stocks by landcover type. Observed carbon pools were modelled as part of a collaboration between Natural Resources Commission of NSW and the Mullion Group (Roberts *et al*., 2022).

### 4.2.4 Outcomes of physical measurement

The model estimated 442 tonnes of sediment were filtered by the riparian land assets in 2013, and that 654 tonnes of sediment would be filtered in the optimal 2023 model, an increase of 212 tonnes per year of sediment filtered (Table 2). The majority of filtration services were supplied by Cumberland Red Gum Riverflat Forest (150 tonnes) and Grass (103 tonnes). The greatest increase in total sediment filtration under the 2023 optimal scenario was also found in Cumberland Red Gum Riverflat Forest, increasing by 141 tonnes. Shale-Sandstone Transition Forest had the highest improvement in sediment filtration per unit area (1.1 t/ha).

The riparian land assets in the optimal 2023 scenario (78,956 tonnes) stored 30,161 tonnes more carbon than the 2013 model (48,795 tonnes), an average increase 81.8 t/ha (Table 3). The greatest increase was in vegetation in the Cumberland Red Gum Riverflat Forest (9,199 tonnes), the greatest average per hectare increase was in the Coastal Valleys Swamp Oak Riparian Forest (153 t/ha), although this asset covered only 9 hectares. Cumberland Red Gum Riverflat Forest had a low per hectare increase (65 t/ha) but covered a large area (141 ha).

Table 2. Sediment filtered from overland erosion flows, total (t) and average per hectare of ecosystem type (t/ha), for the 2013 scenario and the optimal 2023 scenario, and the difference between them. Total row shows sum of tonnes or mean ($\bar{x}$) of tonnes per hectare where indicated. Source: Produced by the authors.

| Asset type | 2013 t/ha. | t. | 2023 t/ha. | t. | Area ha. | Change t/ha. | t. |
|---|---|---|---|---|---|---|---|
| Cumberland Shale Plains Woodland | 0.9 | 50 | 1.3 | 74 | 58 | 0.4 | 24 |
| Cumberland Red Gum Riverflat Forest | 1.1 | 150 | 1.6 | 223 | 141 | 0.5 | 73 |
| Cumberland Shale-Sandstone Ironbark Forest | 1.3 | 33 | 2.3 | 60 | 26 | 1.1 | 28 |
| Coastal Valleys Swamp Oak Riparian Forest | 0.6 | 5 | 1.1 | 10 | 9 | 0.5 | 5 |
| Sydney Turpentine Ironbark Forest | 0.2 | 2 | 0.3 | 3 | 9 | 0.1 | 1 |
| Grass | 1.5 | 103 | 2.2 | 151 | 68 | 0.7 | 48 |
| Non-vegetated still waterbody | 1.8 | 70 | 2.3 | 90 | 39 | 0.5 | 21 |
| Watercourse | 2.2 | 26 | 3.2 | 39 | 12 | 1.1 | 13 |
| Medium Density Urban Fabric | 0.7 | 4 | 0.7 | 5 | 7 | 0 | 0 |
| Total | $\bar{x}$ = 1.2 | 442 | $\bar{x}$ = 1.7 | 654 | 369 | $\bar{x}$ = 0.6 | 212 |

Table 3. Total carbon stored (t), and average carbon per area (t/ha) per vegetation class in the riparian land assets managed area for the 2013 scenario and the optimal 2023 scenario, and the change between the two. Total row shows sum of tonnes or mean ($\bar{x}$) of tonnes per hectare where indicated. Source: Produced by the authors.

| Asset | 2013 | | 2023 | | Area | Change | |
|---|---|---|---|---|---|---|---|
| | t/ha. | t | t/ha. | t | ha. | t/ha. | t |
| Cumberland Shale Plains Woodland | 69 | 3,987 | 215 | 12,446 | 58 | 146 | 8,458 |
| Cumberland Red Gum Riverflat Forest | 158 | 22,382 | 224 | 31,581 | 141 | 65 | 9,199 |
| Cumberland Shale-Sandstone Ironbark Forest | 86 | 2,230 | 215 | 5,590 | 26 | 129 | 3,360 |
| Coastal Valleys Swamp Oak Riparian Forest | 56 | 516 | 210 | 1,916 | 9 | 153 | 1,400 |
| Sydney Turpentine Ironbark Forest | 60 | 535 | 192 | 1,713 | 9 | 132 | 1,178 |
| Grass | 143 | 9,744 | 179 | 12,220 | 68 | 36 | 2,476 |
| Non-vegetated still waterbody | 178 | 6,955 | 242 | 9,455 | 39 | 64 | 2,501 |
| Watercourse | 161 | 1,935 | 267 | 3,201 | 12 | 106 | 1,266 |
| Medium Density Urban Fabric | 79 | 511 | 129 | 835 | 7 | 50 | 323 |
| Total | $\bar{x}$ = 132 | 48,795 | $\bar{x}$ = 214 | 78,956 | 369 | $\bar{x}$ = 82 | 30,161 |

### 4.3 Monetary valuation

### 4.3.1 Sediment filtration

The environmental income from sediment filtration is an estimate of the avoided cost from all sediments filtered by the riparian land assets. Sediment filtration services for 2013 are valued at AU$110,456 using the avoided cost of sediment removal from stormwater infrastructure reported by Perth NRM (Scallan, 2021).

Costs of sediment removal vary widely depending on the asset type, its access or location and type of work required. A wide range of costs are presented in the Perth NRM report, from $250/tonne to $5,143/tonne. We used the lowest cost reported ($250/tonne) to achieve a conservative estimate of value for sediment filtration services.

Present asset values were obtained by discounting the value of future flows of sediment filtration using a private discount rate of 7% to reflect the private nature of the benefit (accruing to business), noting that this private benefit specifically relates to the cost savings provided by natural filtration, rather than the public benefit of improved water quality which accrues to the environment and society. This discount rate is also in line with NSW Treasury recommendations for government cost-benefit analysis (Treasury NSW, 2017). Flows were discounted over 100 years as recommended by the SEEA-EA for assets expected to persist in the long run.

Table 4 shows the estimated value of sediment filtration services for the range of riparian vegetation types present in the catchment. Values for 2013 are compared to values under the optimal 2023 scenario showing an additional $53,057 in service could be generated by improving the sediment filtration capacity of vegetation.

### 4.3.2 Climate regulation

Asset values can be derived by valuing standing carbon using either market values or the social cost of carbon. Stored carbon is valued at Australia's Emissions Reduction Fund (ERF) market price at the time (October 2022), which corresponds to a fair value measure in accounting practice. The social cost of carbon (SCC) is an estimate of the social harm resulting from every additional tonne of $CO_2$-e release to the atmosphere and is typically a lot higher than carbon market values based on abatement costs. SCC is a function of climate system models, the choice of discount rates and the expected damage of rising temperatures on human health, agriculture and sea level rise. The US government's recommended SCC of US$51 (AU$73) per tonne $CO_2$-e is set to increase to US$190 (AU$274) per tonne as a result of updated research (Rennert and Prest, 2022; Rennert et al., 2022).

Table 4 also presents the estimated value of carbon stocks for the range of riparian vegetation types. We applied the more conservative market of AU$37/tonne $CO_2$-e to obtain a standing carbon value of AU$6.6 million for 2013. Values for 2013 are compared to values under the optimal 2023 scenario, showing a potential increase in asset value of $4.0 million, should the land assets (vegetation) be restored to optimum condition. Using the initial (AU$73) and proposed (AU$274) US recommendations for SCC, the value of the carbon stored in the riparian land assets could reach AU$13 million to AU$49 million.

Table 4 - Monetary natural capital account for riparian land assets in 2013 and under optimal scenario. Source: Produced by the authors.

| Asset | Sediment filtration (AU$) | | | Global climate regulation - carbon storage (AU$) | | |
| --- | --- | --- | --- | --- | --- | --- |
| | 2013 | 2023 | Change | 2013 | 2023 | Change |
| Cumberland Shale Plains Woodland | $12,378 | $18,485 | $6,108 | $541,449 | $1,690,017 | $1,148,568 |
| Cumberland Red Gum Riverflat Forest | $37,500 | $55,650 | $18,150 | $3,039,241 | $4,288,323 | $1,249,082 |
| Cumberland Shale-Sandstone Ironbark Forest | $8,150 | $15,085 | $6,935 | $302,839 | $759,127 | $456,288 |
| Coastal Valleys Swamp Oak Riparian Forest | $1,260 | $2,458 | $1,198 | $70,016 | $260,110 | $190,094 |
| Sydney Turpentine Ironbark Forest | $373 | $642 | $269 | $72,704 | $232,639 | $159,934 |
| Grass | $25,744 | $37,822 | $12,078 | $1,323,104 | $1,659,310 | $336,206 |
| Non-vegetated still waterbody | $17,458 | $22,595 | $5,138 | $944,346 | $1,283,886 | $339,540 |
| Watercourse | $6,485 | $9,658 | $3,173 | $262,779 | $434,709 | $171,930 |
| Medium Density Urban Fabric | $1,110 | $1,119 | $9 | $69,439 | $113,333 | $43,895 |
| Total | $110,456 | $163,513 | $53,057 | $6,625,917 | $10,721,454 | $4,095,537 |

The ability to estimate the physical flow of ecosystem services and associated monetary values for different asset types provides an avenue for reporting on change in asset performance over time. The approach would be useful for aligning measures of natural asset performance with standard measures of financial performance, including return on investment for example. The approach could also be used to inform internal asset management decisions, by comparing the outcomes of various investment scenarios across an asset portfolio. In turn, better informed asset management decisions that take into account the contribution of environmental resources can help demonstrate responsible stewardship over those resources.

## 5. Findings and discussion

The purpose of this study was to explore whether natural capital concepts can be incorporated into existing financial accounting practices. Specifically, we asked: Can natural capital be recognised in the financial statements? And if so, what methods can be used in a water utility to recognise natural capital on the balance sheet? Following collaboration with Sydney Water, three possible alternatives are proposed for the financial recognition of the elements of their natural capital (riparian vegetation and carbon storage) identified in this study. They are:

1. Balance Sheet Item
2. Notes to the Financial Statements
3. Voluntary Disclosures

These were presented to Sydney Water as a portfolio of disclosure alternatives, and can also be applied to other water utilities that have similar stormwater management responsibilities, provided they have the necessary resources to undertake the abovementioned measurement techniques.

## Alternative 1: Balance sheet item

In order to determine whether natural capital can be recognised on the balance sheet, we must first consider whether it meets the definition of an asset. An asset is defined as "a present economic resource controlled by the entity as a result of past events. An economic resource is a right that has the potential to produce economic benefits" (AASB Conceptual Framework paras 4.3 and 4.4). These rights take many forms (AASB Conceptual Framework para 4.6), including the provision of economic benefits by enabling the entity to avoid cash outflows through the provision of services (AASB Conceptual Framework para 4.16(c)(i)). The right must have both the potential to produce economic benefits beyond those available to other parties and be controlled by the entity (para 4.9). It does not need to be certain or even likely that the right will produce economic benefits, it is only necessary that the right exists (AASB Conceptual Framework para 4.14).

Based on the definition and requirements set by the AASB, it appears that the riparian vegetation under the control of the water utility meets the criteria of an asset, due to the water filtration services the vegetation provides. These services provide economic benefits in the form of avoided cash outflows, and assuming that no other entity has an obligation to undertake sediment removal in the region, these benefits are not available to other parties. Further, the economic benefits can be controlled by the water utility through their ongoing management of the riparian vegetation. Therefore, we believe the riparian vegetation appears to meet the definition of an asset as per the requirements set out in the AASB Conceptual Framework (2022b), which also suggests that a failure to recognise such items limits the usefulness of the reports, noting that:

> *Not recognising an item that meets the definition of one of the elements makes the statement of financial position and the statement(s) of financial performance less complete and can exclude useful information from financial statements* (AASB Conceptual Framework para 5.7).

The line items that could be included in a balance sheet are set out AASB 101 (para 54), and include, for example, items such as 'Non-current receivables'; 'Property, plant and equipment'; 'Right-of-use assets'; and 'intangible assets'. According to AASB 101 (para 55) an entity shall present additional line items when it is relevant to an understanding of the entity's financial position, thus Sydney Water may present an additional line item of 'natural capital' (or similar), which would allow it to group together additional natural capital assets, as they are identified (as per the definition provided above) and quantified in the future. Therefore, in answer to the first research question, it appears that the water utility may include riparian vegetation as a non-current asset under an additional line item of 'natural capital' on the balance sheet.

The second research question explores what methods can be used in a water utility to recognise natural capital on the balance sheet. As noted previously, the asset value is measured using the value in use method. The present value was calculated by discounting future flows of sediment filtration (conservatively $250/tonne at 442 tonnes per annum) using a private discount rate of 7% to reflect the private nature of the benefit (accruing to the business). This discount rate is also in line with NSW Treasury recommendations for government cost-benefit analysis. Flows were discounted over 100 years as recommended by the SEEA-EA for assets expected to persist in the long run.

This asset valuation requires some estimates and assumptions, and it is noted in the AASB Conceptual Framework (para 2.19) that:

> *The use of reasonable estimates is an essential part of the preparation of financial information and does not undermine the usefulness of the information if the estimates are clearly and accurately described and explained. Even a high level of measurement uncertainty does not necessarily prevent such an estimate from providing useful information.*

The AASB accepts that there may be trade-offs between characteristics of information such as 'relevance' and a 'faithful representation' in order to provide information that is useful for decision-making (AASB Conceptual Framework para 2.22). For example, relevant information may be based on uncertain estimates. It is noted that if the level of measurement uncertainty involved in making an estimate is so high that it is questionable whether the estimate would provide a sufficiently faithful representation, the most useful information may be the uncertain estimate, accompanied by a description of the estimate and the associated uncertainties (para2.22).

### Alternative 2: Notes to the Financial Statements

The notes to the financial statements may provide information that is not provided elsewhere in the financial statements, but that Sydney Water deems relevant to understanding the financial statements or supporting information for items presented in the financial statements (AASB 101 para 114(c)(iii)). The notes may also include information about items that meet the definition of an asset, or other elements, but have not been included in the financial statements (AASB Conceptual Framework, paras 3.3(c)(iii) and 5.6).

The recognition of riparian vegetation as an asset will depend on the extent of Sydney Water's obligation to remove the sediment (and the corresponding rights to the benefits that the riparian vegetation provides). For the purposes of simplicity, in this study the valuation figure is based on the assumption that all of the benefits, in the form of cost savings for water filtration obligations, accrue to the agency. Assuming that the riparian vegetation meets the definition of an asset due to the rights Sydney Water has to the economic benefits that it provides, according to the AASB (Conceptual Framework para 5.11):

> *Even if an item meeting the definition of an asset or liability is not recognised, an entity may need to provide information about that item in the notes. It is important to consider how to make such information sufficiently visible to compensate for the item's absence from the structured summary provided by the statement of financial position and, if applicable, the statement(s) of financial performance.*

This suggests that information regarding natural capital assets that provide economic benefits in the form of avoided cash outflows that accrue to the entity should be provided in the notes, at minimum.

### Alternative 3. Voluntary disclosures

In the absence of a more robust level of disclosure in the financial statements, water utilities such as Sydney Water may choose to provide voluntary disclosures regarding their management of natural capital. Voluntary disclosures in the Annual Report may be in the form of a narrative, or in the form of an Environmental Profit and Loss Statement (Table 5) and Natural Capital Balance Sheet (Table 7).

Currently, there is no requirement for natural capital accounting disclosures to meet the requirements set by the AASB. It should be noted that Tables 5 and 7 are incomplete due to insufficient data, and thus are provided for illustrative purposes only. However, they do demonstrate how organisations may utilise data collected under the SEEA-EA framework and present them in a manner that is consistent with, and understandable by users of, external presentation of financial information.

The Environmental Profit and Loss Statement would be accompanied by Note 1, and the Natural Capital Balance sheet accompanied by Note 2 (the contents of the notes is subject to change, depending on the methods used). Notes 1 and 2 draw from other sections of this paper.

Table 5. Environmental profit and loss statement. Source: Produced by the authors.

| | 2023 | | | | 2013 | | | |
|---|---|---|---|---|---|---|---|---|
| | Measure | Value to business | Value to society | Total | Measure | Value to business | Value to society | Total |
| Environmental income/(loss) | | | | | | | | |
| Sediment filtration (tonnes) *(Note 1)* | xxxx | xxxx | | xxxx | 442 | $110,456 | | $110,456 |
| Carbon sequestration (tonnes) *(Note 2)* | xxxx | | xxxx | xxxx | xxxx | | xxxx | xxxx |
| Total environmental income/(loss) | | xxxx | xxxx | xxxx | | $110,456 | xxxx | $110,456 |

### Note 1

The Rouse Hill riparian land assets consist of natural lands and vegetation located in the riparian zone and managed for the purpose of reducing the sediment load of stormwater before they enter channels and streams. The quantity of sediments filtered by the riparian land assets is modelled using the InVEST (Integrated Value of Ecosystem Services and Tradeoffs) platform developed by Stanford University's Natural Capital Project and Modelled Hillslope Erosion for NSW data on topography, rainfall, land cover management and soil properties.

The environmental income from sediment filtration is an estimate of the avoided cost from all sediments filtered by the riparian land assets. Sediment filtration is valued at the avoided cost of sediment removal from stormwater infrastructure reported by Perth NRM: $250/tonne (see Monetary valuation section for further details).

The following table shows the physical quantity of sediments and the associated value of the filtration services for each asset type presented in the Rouse Hill riparian vegetation (Table 6).

Table 6. Physical quantity of sediments captured by the Rouse Hill riparian vegetation. Source: Produced by the authors.

| Asset | Tonnes of sediment | AU$ |
|---|---|---|
| Cumberland Shale Plains Woodland | 50 | $12,378 |
| Cumberland Red Gum Riverflat Forest | 150 | $37,500 |
| Cumberland Shale-Sandstone Ironbark Forest | 33 | $8,150 |
| Coastal Valleys Swamp Oak Riparian Forest | 5 | $1,260 |
| Sydney Turpentine Ironbark Forest | 1 | $373 |
| Grass | 103 | $25,744 |
| Non-vegetated still waterbody | 70 | $17,458 |
| Watercourse | 26 | $6,485 |
| Medium Density Urban Fabric | 4 | $1,111 |
| Total | 442 | $110,456 |

Table 7. Natural Capital Balance Sheet. Source: Produced by the authors.

| | 2023 | | | | 2013 | | | |
|---|---|---|---|---|---|---|---|---|
| | Measure | Value to business | Value to society | Total | Measure | Value to business | Value to society | Total |
| Sediment filtration (tonnes), Notes 1 | xxxx | xxxx | | xxxx | 441.6 | $1,686,207 | | $1,686,207 |
| Carbon storage (tonnes), Notes 2 | xxxx | | xxxx | xxxx | 179,059 | | $6,625,199 | $6,625,199 |
| Total natural capital assets | | xxxx | xxxx | xxxx | | $1,686,207 | $6,625,199 | $8,311,406 |

## Note 2

Vegetation helps regulate the global climate by removing carbon from the atmosphere and storing it as living or dead biomass. Carbon storage and sequestration are separate processes that contribute to climate regulation. Estimates of carbon sequestration over an accounting period are obtained by the difference in carbon stocks between two time periods.

Biomass carbon in the riparian land assets was modelled using the InVEST (Integrated Valuation of Ecosystem Services and Tradeoffs) platform developed by Stanford University's Natural Capital Project (Sharp *et al.*, 2018), data from the NSW Forest Carbon Stock 1990-2020, the NSW State Vegetation Type Map (DPE, 2022). Three carbon pools are included in the carbon stock assessment: aboveground, belowground and dead woody debris. Litter and soil carbon are excluded so the assessment is most likely an underestimate of the total carbon present on site.

The stock of carbon stored on site is measured at fair value using the Australian Emissions Reduction Fund market price of AU\$37/tonne of $CO_2$-e (January 2023). Modelled biomass carbon stocks convert to 3.67 tonnes of $CO_2$-e. The contribution of sediment filtration services to the asset value is measured using the value in use method. Present asset values were obtained from the discounted rate of future cash flows from sediment filtration using a private discount rate of 7% to reflect the private nature of the benefit (accruing to business), noting that this private benefit specifically relates to the cost savings provided by natural filtration, rather than the public benefit of improved water quality which accrues to the environment and society. This discount rate is also in line with NSW Treasury recommendations for government cost-benefit analysis. Flows were discounted over 100 years as recommended by SEEA-EA for assets expected to persist in the long run. The following table shows the physical quantity of biomass carbon and the associated value of the filtration services for each natural asset type present in the Rouse Hill riparian land assets (Table 8).

Table 8. Physical quantity of biomass carbon in 2013 and the associated value of the carbon storage services for each asset type present in the Rouse Hill riparian land assets. Source: Produced by the authors.

| Asset | Tonnes biomass carbon | Tonnes CO-e | AU$ |
| --- | --- | --- | --- |
| Cumberland Shale Plains Woodland | 3,987 | 14,634 | $541,449 |
| Cumberland Red Gum Riverflat Forest | 22,382 | 82,142 | $3,039,241 |
| Cumberland Shale-Sandstone Ironbark Forest | 2,230 | 8,185 | $302,839 |
| Coastal Valleys Swamp Oak Riparian Forest | 516 | 1,892 | $70,016 |
| Sydney Turpentine Ironbark Forest | 535 | 1,965 | $72,704 |
| Grass | 9,744 | 35,760 | $1,323,104 |
| Non-vegetated still waterbody | 6,954 | 25,523 | $944,346 |
| Watercourse | 1,935 | 7,102 | $262,779 |
| Medium Density Urban Fabric | 511 | 1,877 | $69,439 |
| Total | 48,795 | 179,079 | $6,625,917 |

## 6. Conclusions

The primary focus of this study has been to explore how economic and environmental accounting methodologies can assist in providing financial recognition of publicly owned natural capital assets. This research showed that many SEEA concepts and data organisation principles are useful to guide the recognition of natural capital in financial statements. Environmental accounting can provide multiple benefits including increasing the visibility of natural assets, making it easier to link to business planning and demonstrate environmental stewardship. The value of the contribution made to a water utility's business objectives (water filtration) and to broader community wellbeing (carbon sequestration) were also distinguished.

Current accounting standards can allow the recognition of environmental income or assets, as conceptually the economic resource is the set of rights that produce the economic benefit, not the physical object (AASB Conceptual Framework paras 4.9 & 4.12). This research described how a water utility managing riparian lands for the benefit of stormwater filtration and carbon sequestration could disclose changes in these assets either through voluntary disclosures, in notes to the financial statements or as balance sheet items, and that failing to do so may limit the usability of the financial statements, under the Accounting Standards. Cost savings resulting from ecosystem services could give rise to the recognition of a natural asset, provided that the organisation can demonstrate that the economic benefits (cost savings) are controlled by the organisation and not available to other parties (AASB Conceptual Framework para 4.9).

However, limitations with current accounting practices mean that they cannot incorporate non-productive environmental income, for example from carbon sequestration or for biodiversity improvements. These limitations have implications for public agencies, where monetary valuation or any other quantification of stocks and flows of natural capital is context dependent. The critical context here is the responsibility given officially to the public agency in relation to natural capital. For water utilities with environmental accountabilities, greater visibility and recognition of natural assets in decision making can be improved using this methodology. This in turn enables the agency to discharge their environmental accountabilities by demonstrating their stewardship over the resources under their control.

The natural capital assessment conducted in this study pointed to novel applications for ecosystem service science in environmental management accounting, and emphasises the importance of interdisciplinary research. More specifically, the research tested the assessment of natural asset performance through modelling change in ecosystem service flows over time and across different asset types. To this end, ecosystem service modelling using the InVEST platform was successful at estimating the value of investment in natural capital. Further research extending to different natural settings and different institutional arrangements would improve our understanding of the approach, areas of potential applications and technological limits.

## Acknowledgements

The authors would like to acknowledge the NSW Department of Planning and Environment for their financial support of this research project. We would also like to extend our gratitude to Justine Trounce for her assistance in managing the project.